# Anomalous dynamics of cell migration

**Peter Dieterich***, **Rainer Klages¶, Roland Preuss§, and Albrecht Schwab#**

———————

* Institut für Physiologie, Medizinische Fakultät Carl Gustav Carus, Fetscherstr. 74, D-01307 Dresden, Germany

¶ School of Mathematical Sciences, Queen Mary, University of London, Mile End Road, London E1 4NS, UK

§ Max-Planck-Institut für Plasmaphysik, EURATOM Association, D-85748 Garching, Germany

# Institut für Physiologie II, Robert-Koch-Str. 27b, D-48149 Münster, Germany

**Corresponding author**

Dr. Peter Dieterich, Institut für Physiologie, Medizinische Fakultät Carl Gustav Carus,

Fetscherstr. 74, D-01307 Dresden, Germany

Tel.: ++49 351 458 6042, Fax: ++49 351 458 6301

email: peter.dieterich@tu-dresden.de

**ABSTRACT**

**Cell movement, for example during embryogenesis or tumor metastasis, is a complex dynamical process resulting from an intricate interplay of multiple components of the cellular migration machinery. At first sight, the paths of migrating cells resemble those of thermally driven Brownian particles. However, cell migration is an active biological process putting a characterization in terms of normal Brownian motion into question. By analyzing the trajectories of wildtype and mutated epithelial (MDCK-F) cells we show experimentally that anomalous dynamics characterizes cell migration. A superdiffusive increase of the mean squared displacement, non-Gaussian spatial probability distributions, and power-law decays of the velocity autocorrelations are the basis for this interpretation. Almost all results can be explained with a fractional Klein-Kramers equation allowing the quantitative classification of cell migration by a few parameters. Thereby it discloses the influence and relative importance of individual components of the cellular migration apparatus to the behavior of the cell as a whole.**

**INTRODUCTION**

Nearly all cells in the human body are mobile at a given time during their life cycle. Embryogenesis, wound-healing, immune defense, and the formation of tumor metastases are well-known phenomena that rely on cell migration. Extensive experimental work revealed a precise spatial and temporal coordination of multiple components of the cellular migration machinery such as the actin cytoskeleton, cell-substrate and cell-cell interactions, and the activity of ion channels and transporters[1-4]. These findings are the basis for detailed molecular models representing different microscopic aspects of the process of cell migration like the protrusion of the leading edge of the lamellipodium, or actin dynamics[5]. Mathematical continuum models, on the other hand, focus on collective properties of the entire cell in order to explain requirements for the onset of motion and some typical features of cell motility[6]. These models are usually limited to small spatiotemporal scales. Therefore they provide little information about how the integration of protrusion of the lamellipodium, retraction of the rear part, and force transduction onto the extracellular matrix leads to the sustained long-term movement of the entire cell. This process is characterized by alternating phases of directed migration, changes of direction, and polarization. The coordinated interaction of these phases suggests the existence of intermittency[7] and of strong spatiotemporal correlations. It is therefore an important question whether the long-term movement of the entire cell can still be understood as a simple diffusive behavior like usual Brownian motion[8,9] or whether more advanced concepts of dynamic modeling have to be applied[10,11].

**RESULTS AND DISCUSSION**

We performed migration experiments and analyzed the trajectories of two migrating transformed renal epithelial MDCK-F cell strains: wildtype (NHE$^+$) and NHE-deficient (NHE$^-$) cells[12]. The cells were observed for up to 1000 min. Fig. 1a depicts the contours and the path of a migrating MDCK-F NHE$^+$ cell monitored for 480 min. At first sight, the cell's trajectory resembles those of normal Brownian particles. Brownian motion in terms of the Ornstein-Uhlenbeck process[13,14] is characterized by a mean squared displacement *msd* (for definition see Eq. 1) proportional $\sim t^2$ at short times corresponding to ballistic motion and $\sim t$ for long time intervals designating normal diffusion. Our experimental data show that both types of MDCK-F cells behave differently. Consistent with earlier observations MDCK-F NHE$^-$ cells move less efficiently than NHE$^+$ cells[12,15,16] resulting in a reduced *msd* for all times. As displayed in Fig. 1b, their mean squared displacement *msd* exhibits a crossover between three different dynamical regimes. For short times (< 4 min, phase I), the increase of the *msd* differs from a ballistic $t^2$ scaling. The logarithmic derivative of the *msd* (Eq. 2) shown in Fig. 1c characterizes this first region with an exponent $\beta(t)$ below ~ 1.8. In the subsequent intermediate phase II (up to ~ 20 min) the *msd* reaches its strongest increase with a maximum exponent of $\beta(t) \sim 1.8$. When the cell has roughly moved beyond a squared distance larger than its mean squared radius $r$ (indicated by arrows in Fig. 1b), the exponent $\beta(t)$ of the *msd* gradually decreases below 1.5.

We next extracted the probability that the cells reach a given position $x$ at time $t$ from the experimental data. This corresponds to the temporal development of the spatial probability distribution function $p(x,t)$ delivering information beyond the *msd*. Figs. 2a and 2b reveal the existence of non-Gaussian $p(x,t)$ distributions for different points in time. The transition from a peaked distribution at short times $t = 1$ min to rather broad distributions at long times $t = 120$ min and $t = 480$ min in Figs. 2a and 2b suggests again the existence of distinct dynamical processes acting on different time scales. The shape of the distributions can be quantified by calculating the kurtosis according to Eq. 5 which is displayed as a function of time in Fig. 2c. The kurtosis rapidly decays from values around 7 - 9 to reach a constant value of about ~ 2.3 for both cell types in the long time limit. Such a behavior implies a transition of $p(x,t)$ from a peaked to a flat form also visible from Figs. 2a and 2b. The time-dependent deviation of the kurtosis from the value of *3* which would correspond to a Gaussian distribution (as e.g. for the Ornstein-Uhlenbeck process) is another strong manifestation of the anomalous nature of cell migration.

To gain further insight into the origin of the anomalous dynamics we calculated the velocity autocorrelation function $v_{ac}(t)$ (defined in Eq. 4) that characterizes the correlation of the velocity (Eq. 3) at time $t$ with its value at time $t = 0$ (Fig. 3). As for the *msd* in Fig. 1b, the double-logarithmic plot displays transitions between three regimes characterized by different time scales for both cell types. After a pronounced dip at short times, the autocorrelation function shows a gradual transition during intermediate times to a power-law like decay at long times. In contrast, the Ornstein-Uhlenbeck realization of Brownian motion would imply a purely exponential decay of the velocity autocorrelation function for all times. The non-trivial behavior of the velocity autocorrelation function stresses the existence of long-range correlations in time of the underlying dynamical process of cell migration. We would like to emphasize at this point that no chemotactic or other physical gradients were imposed on the cells examined in our study. Thus, our analysis shows that the so-called 'random migration' does actually not proceed as randomly as one might expect.

In order to generalize the interpretation of our data we sought for an integrative mathematical model. The experimental results posed extensive constraints on the choice of such a model. A full theory should generate a power-law behavior of the *msd* including transitions between different time scales, non-Gaussian probability distributions, and a velocity autocorrelation function with a power-law decay for long times. Conventional

Langevin, Fokker-Planck, or Klein-Kramers equations[17] provide descriptions of ordinary Brownian motion (e.g. the Ornstein-Uhlenbeck process[13,14]) which do not match these constraints. The anomalous nature of cell migration demands for the inclusion of temporal memory in the above equations which can be achieved by introducing fractional derivatives[18-21]. We therefore modeled our *msd* data with the fractional Klein-Kramers equation in Eq. 6[18] which includes the Orstein-Uhlenbeck process as special case ($\alpha = 1$). The inclusion of an uncorrelated noise term delivers the formula for the *msd* as given in Eq. 10. The continuous blue and orange lines in Fig. 1b represent the resulting fits with parameter values as given in Table 1a for MDCK-F NHE$^+$ and NHE$^-$ cells. There is excellent agreement of data and model for all times. The shallow initial slope of the *msd* curve is due to the estimated noise level $\eta \sim$ 0.78 μm and 0.47 μm for NHE$^+$ and NHE$^-$ cells, respectively. The values of $\eta$ are much larger than the measurement uncertainty of $\sigma_{pos} \sim 0.1$ μm showing the influence of 'biological noise' generated by lamellipodial activity. The time scale $\tau_\alpha = (1/\gamma_\alpha)^{1/\alpha}$ (Table 1a and Eqs. 6-8) characterizes the transition of the Mittag-Leffler function from stretched exponential to power-law behavior for $t >> \tau_\alpha$. The resulting time scales $\tau_\alpha = 18.7$ min for NHE$^+$ and $\tau_\alpha = 15.2$ min for NHE$^-$ are comparable with the times at which the cells cross their mean squared radii (458 μm$^2$ and 211 μm$^2$ for NHE$^+$ and NHE$^-$ cell, respectively). For larger times, the *msd* shows a transition to a power-law $\sim t^{2-\alpha}$ (see Eq. 8) thus describing superdiffusion $\sim t^{1.25}$ for NHE$^+$ and $\sim t^{1.28}$ for NHE$^-$ cells. The enhanced thermal velocity $v_{th}^2$ and the slightly reduced value of $\gamma_\alpha$ generate a diffusion coefficient for NHE$^+$ cells that is twice as large as that for NHE$^-$ cells. Thus, our model quantitatively confirms the importance of the Na$^+$/H$^+$ exchanger for directed migration[15,16]. The logarithmic derivative of the model Eq. 10 in Fig. 1c emphasizes the influence of the noise term for short times generating the deviation from a ballistic initial increase of the *msd* in agreement with the experimental data. Without the noise term ($\eta = 0$) $\beta(t)$ would be 2 for short times. In addition, the figure shows a continuous transition of $\beta(t)$ to the estimated exponent 2-$\alpha$ (Table 1a) for long times.

In order to show that the predictions of the fractional Klein-Kramers equation correspond more closely to the experimental data than those from simpler dynamical models we included a quantitative analysis of the Ornstein-Uhlenbeck process[13,14,17]. The application of Eq. 11 to the experimental *msd* data delivers the Ornstein-Uhlenbeck parameters in Table 1b. At first sight the double logarithmic plot in Fig. 1b shows only small differences between the fractional model and the conventional Ornstein-Uhlenbeck process. However, the logarithmic derivative shown in Fig. 1c is clearly in favor of the fractional Klein-Kramers equation, especially in phases II and III. For longer times, the Ornstein-Uhlenbeck process shows a too fast decay of the exponent $\beta(t)$ towards values near one corresponding to normal diffusion.

The probability distribution $p(x,t)$ of the fractional Klein-Kramers equation is only known in the limit of long times ($t \gg \tau_\alpha$) given by the solution of the corresponding fractional diffusion equation[19] in terms of Fox functions (Eq. 9). Figs. 2a and 2b compare these model solutions (using the parameters of Table 1a) with data for times $t = 120$ and 480 min for MDCK-F NHE$^+$ and NHE$^-$ cells, respectively. There is a good agreement between data and model for longer times. The rather peaked solutions in Fig. 2a and 2b for the shortest time $t = 1$ min cannot be explained with these functions. For comparison we have also added the Gaussian probability distributions of the Ornstein-Uhlenbeck process (green lines in Figs. 2a and 2b) which can neither explain the peaked short time nor the long time behavior. Fig. 2c illustrates that the kurtosis of the fractional Klein-Kramers equation deviates for short times, too. However, for longer times data and theoretical kurtosis converge in agreement with the observation of $p(x,t)$ and the Fox functions in Figs. 2a and 2b towards a value around 2.3. In contrast, the Ornstein-Uhlenbeck process has a constant value of 3 over the entire time range.

7 - 20

Finally, in Fig. 3 we compare the velocity autocorrelation function of the fractional Klein-Kramers equation with the cell data. On the double logarithmic plot the Mittag-Leffler function given by Eq. 7 (using the same parameters of Table 1a from the *msd* fit) nicely interpolates between the experimental data values again showing the crossover between stretched exponential and power-law behavior within the same time scales as the *msd* in Fig. 1b. In contrast, the velocity autocorrelation function of the Ornstein-Uhlenbeck process (green lines in Fig. 3) decays too fast ~ $exp(-\gamma_1 t)$ (see Eq. 7) thereby missing the long-range correlations of the fractional model. The obvious deviations of the first data point at $t = 1$ min for both cell types from the fractional Klein-Kramers model can be explained with the contribution of the uncorrelated 'biological noise' to the velocity autocorrelation function as given by Eq. 12. Using the values of $\eta$ from Table 1a the differences between the model solutions and the data points at $t = 1$ min are estimated as 1.23 $\mu m^2/min^2$ and 0.45 $\mu m^2/min^2$ for MDCK-F $NHE^+$ and $NHE^-$ cells, respectively. This agrees quite well with the observed differences in Fig. 3 and is also the case for the corrections at $t = 0$ min which are not visible in the double logarithmic plots.

In summary, we have shown that a variety of anomalous dynamical properties characterizes the migration process of MDCK-F cells. In all these quantities we observe a crossover between anomalous dynamics on different time scales which reminds of intermittent behavior as claimed to be important for optimal search strategies of foraging animals[7]. The fractional Klein-Kramers equation amended by an uncorrelated noise term models the *msd* and velocity autocorrelation function for all times as well as the long time dynamics of the probability distribution *p(x,t)*. Thus, our approach offers a theoretical framework which allows the classification of the dynamics of cell migration with a few physical parameters that can be calculated from the cells' trajectories. These can be compared for different cell types and under different experimental conditions. We probed our model by comparing migration of wildtype and NHE-deficient MDCK-F cells. The defect in directional migration induced by NHE-deficiency[12,15] can clearly be detected by analyzing the dynamics of MDCK-F cell migration. However, it is remarkable that the general pattern of anomalous dynamics is not changed by NHE-deficiency. This observation indicates that the dynamics of cell migration is organized on a level of complexity that is above that of individual components of the cellular migration machinery. On the other hand, our model may allow the identification of those components of the cellular migration apparatus that govern the long-term behavior of migrating cells.

## MATERIALS AND METHODS

### Cell culture

Experiments were carried out on transformed Madin-Darby canine kidney (MDCK-F) cells and on NHE-deficient MDCK-F cells[12] referred to as $NHE^+$ and $NHE^-$ MDCK-F cells. Cells were kept at 37°C in humidified air containing 5% $CO_2$ and grown in bicarbonate-buffered Minimal Essential Medium (MEM; pH 7.4) with Earle's salts (Biochrom, Berlin, Germany) supplemented with 10% fetal calf serum (Biochrom).

### Migration experiments

Cells were seeded at low density (in order to avoid collisions between cells during the experiment) in tissue culture flasks (Falcon) 1 to 2 days prior to the experiments. They were incubated in HEPES-buffered (20 mmol/l, pH 7.4) MEM supplemented with 10 % fetal calf serum during the course of the experiments. The flasks were placed in a heating chamber (37 °C) on the stage of an inverted microscope equipped with phase contrast optics (Axiovert 40; Zeiss, Oberkochen, Germany). Migration was monitored with a video camera (Hamamatsu, Hersching, Germany) controlled by HiPic software (Hamamatsu). Images were taken in 1 min time intervals during a time range of up to 1000 min. $N = 13$ trajectories of each cell type were used for the analysis consisting of more than 10000 data points for each group.

### Data analysis

Image segmentation was performed with Amira software (Mercury Computer Systems, USA; http://www.amiravis.com/). The outlines of individual cells at each time step were marked throughout the entire image stack and taken for all further processing. In addition, we assessed the accuracy of the segmentation by repeated segmentation. The normally distributed experimental uncertainty amounts to $\sigma_{pos} \sim 0.1$ µm (data not shown).

Quantitative data analysis and calculation of parameters were performed with programs developed by ourselves. The *x*- and *y*-coordinates of the cell centre (µm) were determined as geometric means of equally weighted pixel positions within the cell outlines as function of time. The combined trajectories of a cell population allow the calculation of the mean squared displacement *msd* that describes the mean of the squared distances between a common starting point at time $t_0$ and the actual positions of a cell population at time *t*,

$$msd(t) = \langle [x(t+t_0) - x(t_0)]^2 + [y(t+t_0) - y(t_0)]^2 \rangle \tag{1}$$

where $\langle ... \rangle$ denotes a combined average over all starting times $t_0$ and cell paths. The increase of the *msd* can be quantified by the logarithmic derivative

$$\beta(t) = \frac{d \ln msd(t)}{d \ln t} \tag{2}$$

leading to a time-dependent increase $msd(t) \sim t^{\beta(t)}$.

The velocity of migrating cells in *x*- and *y*-direction ($v_{x/y}(t)$ [µm/min]) was calculated from trajectories as the difference quotient of two cell positions at times $t + \Delta$ and $t$

$$v_x(t) = \frac{x(t+\Delta) - x(t)}{\Delta} \quad \text{and} \quad v_y(t) = \frac{y(t+\Delta) - y(t)}{\Delta} \tag{3}$$

with the observation time interval of $\Delta = 1$ min. These velocities were used to calculate the velocity autocorrelation function

$$v_{ac}(t) = \langle v_x(t+t_0) v_x(t_0) \rangle + \langle v_y(t+t_0) v_y(t_0) \rangle \tag{4}$$

where $\langle ... \rangle$ denotes an average as in Eq. 1.

The position of the cells can also be used to calculate the probability *p(x,y,t)* of finding a cell at position *(x,y)* for time *t*. Because we did not find any correlations between *x* and *y* direction in the velocity autocorrelation function (data not shown) we reduced the discussion to *p(x,t)* given as average of *x* and *y* positions.

The kurtosis is defined as the ratio of moments by

$$\kappa = <x^4(t)>/<x^2(t)>^2. \tag{5}$$

It can be interpreted as shape index of the probability distribution function $p(x,t)$ and takes the value of 3 for Gaussian functions.

**Fractional Klein-Kramers equation**

The anomalous properties of cell dynamics were assessed with the fractional Klein-Kramers equation (FKK) for the probability distribution $P(x, v, t)$ in position $x$, velocity $v$, and time $t$ as proposed by Barkai and Silbey[18] but without external forces:

$$\frac{\partial P(x,v,t)}{\partial t} = -\frac{\partial}{\partial x}[vP(x,v,t)] + \frac{\partial^{1-\alpha}}{\partial t^{1-\alpha}} \gamma_\alpha \left[ \frac{\partial}{\partial v} v + \frac{k_B T}{M} \frac{\partial^2}{\partial v^2} \right] P(x,v,t). \tag{6}$$

$\gamma_\alpha$ denotes the damping term, $k_B$ the Boltzmann constant, $T$ the temperature, $M$ the mass of the particle, and $\alpha$ defines the order of the fractional time derivative of Riemann-Liouville type (see for example[20]). For $\alpha = 1$ the above equation reduces to the ordinary Klein-Kramers equation[17]. The fractional Klein-Kramers equation of Eq. 6 implies that the velocity correlation function is given by the so-called Mittag-Leffler function[22] $E_\alpha$

$$\langle v_x(t) v_{x0} \rangle = v_{th}^2 E_\alpha(-\gamma_\alpha t^\alpha) \quad \Rightarrow \quad v_{ac}(t)_{FKK-2D} = 2v_{th}^2 E_\alpha(-\gamma_\alpha t^\alpha). \tag{7}$$

In the case of the Ornstein-Uhlenbeck limit ($\alpha = 1$) $E_\alpha$ reduces to an exponential decay ~ $exp(-\gamma_1 t)$. The mean squared displacement of the fractional Klein-Kramers equation is represented by the generalized Mittag-Leffler function[22] $E_{\alpha,\beta}$

$$msd(t)_{FKK-1D} = 2v_{th}^2 t^2 E_{\alpha,3}(-\gamma_\alpha t^\alpha) \sim \frac{2 D_\alpha t^{2-\alpha}}{\Gamma(3-\alpha)} \quad \text{for} \quad t \to \infty. \tag{8}$$

The mean thermal velocity $v_{th}^2 = k_B T/M$ is related to the generalized diffusion coefficient $D_\alpha$ by the relation $D_\alpha = v_{th}^2 / \gamma_\alpha$. In the limit of large damping ($\gamma_\alpha \to \infty$) the fractional Klein-Kramers equation reduces to a fractional diffusion equation[19,20] (FDE) where the spatial probability distribution functions $p(x,t)$ are given by a Fox function $H$ [19,20]:

$$p(x,t)_{FDE} = \frac{1}{\sqrt{4\pi D_\alpha t^{2-\alpha}}} H_{12}^{20}\left[\frac{x^2}{4 D_\alpha t^{2-\alpha}} \middle| \begin{array}{cc} (\alpha/2, 2-\alpha) \\ (0,1) & (1/2,1) \end{array}\right]. \qquad (9)$$

Eq. 8 can be extended in order to include uncorrelated noise of variance $\eta^2$ generated by measurement errors[23] or by biological activity e.g. by the fluctuating lamellipodium (t > 0)

$$msd(t)_{FKK-2D+noise} = 4 v_{th}^2 t^2 E_{\alpha,3}(-\gamma_\alpha t^\alpha) + (2\eta)^2. \qquad (10)$$

Eq. 10 reduces to the Ornstein-Uhlenbeck result for $\alpha = 1$ with the noise term:

$$msd(t)_{OU-2D+noise} = \frac{4 v_{th}^2}{\gamma_1^2}\left(\gamma_1 t - 1 + e^{-\gamma_1 t}\right) + (2\eta)^2. \qquad (11)$$

In a similar way, uncorrelated noise also influences the velocity autocorrelation function. The fluctuations $\eta_i$ affect the velocity calculation in Eq. 3 at both positions $x(t_i)$ and $x(t_i+\Delta)$ separated by the measurement interval $\Delta$. Performing the average for the calculation of the velocity autocorrelation function eliminates all linear terms ~ $\eta_i$ while quadratic noise terms deliver a modification of the velocity autocorrelation function at times $t = 0$ and $t = \Delta$

$$v_{ac}(t)_{FKK-2D+noise} = v_{ac}(t)_{FKK-2D} + \frac{4\eta^2}{\Delta^2}\delta(t,0) - \frac{2\eta^2}{\Delta^2}\delta(t,\Delta) \qquad (12)$$

with the Kronecker delta $\delta(t,t') = 1$ for $t = t'$ and 0 elsewhere. All other times are unaffected, if the noise source is uncorrelated.

**Bayesian data analysis**

The parameters of the FKK model in Eq. 6 and their uncertainties were estimated with Bayesian data analysis[24] (for a recent review see[25]) applied to the corresponding two-dimensional mean squared displacement of Eq. 10 (or Eq. 11 for the Ornstein-Uhlenbeck process). Bayesian data analysis offers a logically consistent link between data and models. It allows a reliable estimation of the model parameters taking the uncertainties of the experimental data into account. The expectation value of the *n*-th moment of the parameters $\hat{\theta} = \{\alpha, \gamma_\alpha, v_{th}^2, \eta\}$ in Eq. 10 is given by integrating over the posterior probability function for these parameters $p(\hat{\theta} | data)$. The later is proportional to the product of likelihood $p(data | \hat{\theta})$ and prior function $p(\hat{\theta})$:

$$\langle \theta_i^n \rangle = \frac{\int \prod_{k=1}^{4} d\theta_k \, \theta_i^n \, p(data | \hat{\theta}) \, p(\hat{\theta})}{\int \prod_{k=1}^{4} d\theta_k \, p(data | \hat{\theta}) \, p(\hat{\theta})} \quad . \tag{13}$$

Assuming normally distributed errors, the likelihood function $p(data | \hat{\theta})$ is given by

$$p(data | \hat{\theta}) = \left( \prod_{i=1}^{N} \frac{1}{\sqrt{2\pi\sigma^2(t_i)}} \right) \exp\left\{ -\frac{1}{2} \sum_{i=1}^{N} \left( \frac{data(t_i) - msd(\hat{\theta}, t_i)}{\sigma(t_i)} \right)^2 \right\} \tag{14}$$

where $msd(\hat{\theta}, t_i)$ corresponds to Eq. 10 for the FKK model. A constant prior $p(\hat{\theta})$ was applied due to the lack of in advance information about the values of the parameters. The 4-dimensional integral in Eq. 13 was evaluated numerically by Markov Chain Monte Carlo sampling. Thus, application of Eq. 13 with *n = 1* delivers the expectation values of the four parameters $\alpha$ [1], $\gamma_\alpha$ [1/min$^\alpha$], $v_{th}^2$ [µm$^2$/min$^2$], and $\eta$ [µm] based on the available experimental *msd* data at time points $t_i$. The error $\sigma(t_i)$ was estimated as

$$\sigma(t_i) = data(t_i) / \sqrt{N T_{path} / t_i} \tag{15}$$

where *N* is the number of cell paths. The quotient of path length $T_{path}$ (= 500 min) and actual time $t_i$ estimates the number of more or less independent measurement intervals during the calculation of the combined expectation value in Eq. 1 (also see Qian et al.[26] for the discussion of statistical errors of the *msd*). In addition, Eq. 13 allows the estimation of the uncertainties of the parameter with *n = 2* via

$$\delta\theta_k = \sqrt{\langle \theta_k^2 \rangle - \langle \theta_k \rangle^2} \quad . \tag{16}$$

In addition, we applied this formalism to simulated data of the Ornstein-Uhlenbeck process with a number of data that is comparable to the experiments. These simulations showed that the presented Bayesian data analysis delivers an agreement of estimated and actually used simulation parameters within the uncertainties (data not shown).

**AUTHORS' CONTRIBUTIONS**

P.D. accomplished the main analysis, R.K. focused on the modeling part, R.P. is responsible for the Bayesian data analysis, and A.S. performed the experiments and acquired all experimental data. The interdisciplinary paper was written by all authors.

**ACKNOWLEDGEMENTS**

We thank Eric Lutz for helpful discussions about anomalous dynamics, Andreas Deussen and Hans Oberleithner for fruitful comments, and Sabine Mally for excellent technical assistance. R.K. thanks the MPIPKS Dresden for frequent hospitality. This work was supported by grants from the Deutsche Forschungsgemeinschaft Schw 407/9-3 and Schw 407/10-1 to A.S. and by a grant from the British EPSRC EP/E00492X/1 to R.K.

**FIGURE LEGENDS**

**Fig. 1:** Summary of migration experiments. **a.** shows an overlay of a migrating MDCK-F NHE$^+$ cell with its path covered within 480 min. The cell frequently changes its shape and direction during migration. **b.** presents a double-logarithmic plot of the mean squared displacement (*msd*) as a function of time. Experimental data points for both cell types are symbolized by triangles and circles. Different time scales are marked as phases I, II, and III as discussed in the text. The solid blue and orange lines represent the fit to the *msd* of the fractional Klein-Kramers (FKK) equation including a noise term (Eq. 10). Green lines show the results of the Ornstein-Uhlenbeck (OU) model plus noise (Eq. 11). The corresponding parameters of the theoretical models are given in Table 1. The dashed black lines indicate the uncertainties of the *msd* values according to Eq. 15. **c.** displays the logarithmic derivative *β(t)* (Eq. 2) of the *msd* for MDCK-F NHE$^+$ and NHE$^-$ cells. Data and model curves are marked as in Fig 1b.

**Fig. 2:** Time-dependent development of the spatial probability distribution *p(x,t)*. **a.** and **b.** show the experimental data for NHE$^+$ and NHE$^-$ cells, respectively, at different time points *t* = 1, 120, and 480 min in a semi-logarithmic representation. The continuous blue and orange lines show the solutions of the fractional diffusion equation as given in Eq. 9 with the parameter set obtained by the *msd* fit in Table 1a. Analogously, the green lines depict the Gaussian Ornstein-Uhlenbeck functions. For *t* = 1 min the probability distribution of the data points shows a peaked structure clearly deviating from a Gaussian form. **c.** The kurtosis of the distribution function *p(x,t)* varies as a function of time and saturates around ~2.3 for long times. Being different from the value of *3* (green line, OU), the kurtosis confirms the deviation from a Gaussian (Ornstein-Uhlenbeck) probability distribution. The continuous blue and orange lines (FKK) represent the kurtosis modeled with the fractional Klein-Kramers

equation including a Gaussian noise term.

**Fig. 3:** Decay of the velocity autocorrelation function. The points represent the experimental data for MDCK-F NHE$^+$ (**a**.) and NHE$^-$ (**b**.) cells. The continuous blue and orange curves display the velocity correlation function of the Klein-Kramers equation (FKK) given by the Mittag-Leffler function in Eq. 7 with the parameters of Table 1a. The uncertainties of the fractional Klein-Kramers model estimation are indicated as dashed black lines. The green lines display the exponential decay of the Ornstein-Uhlenbeck process (OU). While the fractional velocity autocorrelation function of the Klein-Kramers equation reliably models the experimental data, the Ornstein-Uhlenbeck process fails to do so.

**Table 1: Parameter estimation.**

a. Fractional Klein-Kramers equation.

| parameter/data | $\alpha$ [1] | $\gamma_\alpha$ [1/min$^\alpha$] | $v_{th}^2$ [µm$^2$/min$^2$] | $\eta$ [µm] | $\tau_\alpha$ [min] | $D_\alpha$ [µm$^2$/min$^{2-\alpha}$] |
|---|---|---|---|---|---|---|
| NHE$^+$ | 0.754 | 0.110 | 0.680 | 0.783 | 18.68 | 6.18 |
|  | ± 0.015 | ± 0.010 | ± 0.015 | ± 0.008 | ± 2.50 | ± 0.58 |
| NHE$^-$ | 0.717 | 0.142 | 0.437 | 0.474 | 15.22 | 3.08 |
|  | ± 0.017 | ± 0.015 | ± 0.012 | ± 0.008 | ± 2.45 | ± 0.34 |

b. Ornstein-Uhlenbeck process.

| parameter/data | $\alpha$ [1] | $\gamma_1$ [1/min] | $v_{th}^2$ [µm$^2$/min$^2$] | $\eta$ [µm] | $\tau_1$ [min] | $D_1$ [µm$^2$/min] |
|---|---|---|---|---|---|---|
| NHE$^+$ | 1 | 0.028 | 0.549 | 0.830 | 36.22 | 19.88 |
|  | - | ± 0.001 | ± 0.019 | ± 0.007 | ± 0.70 | ± 0.95 |
| NHE$^-$ | 1 | 0.029 | 0.339 | 0.522 | 34.09 | 11.57 |
|  | - | 0.001 | ± 0.012 | ± 0.005 | 0.70 | ± 0.58 |

Parameters and their uncertainties (SD) were estimated with Bayesian data analysis applied to (a.) the experimental data and the *msd* of the fractional Klein-Kramers equation (FKK) supplemented with an uncorrelated noise term as given in Eq. 10, or (b.) the Ornstein-Uhlenbeck (OU) result in Eq. 11, for both cell types MDCK-F NHE$^+$ and NHE$^-$. A reliable parameter estimation can be performed for both models. However, the predictions of the fractional Klein-Kramers equation taking into account anomalous dynamics are superior to the Ornstein-Uhlenbeck process as shown in Figs. 1-3. The missing anomalous features of the Ornstein-Uhlenbeck process imply the modification of the parameters.

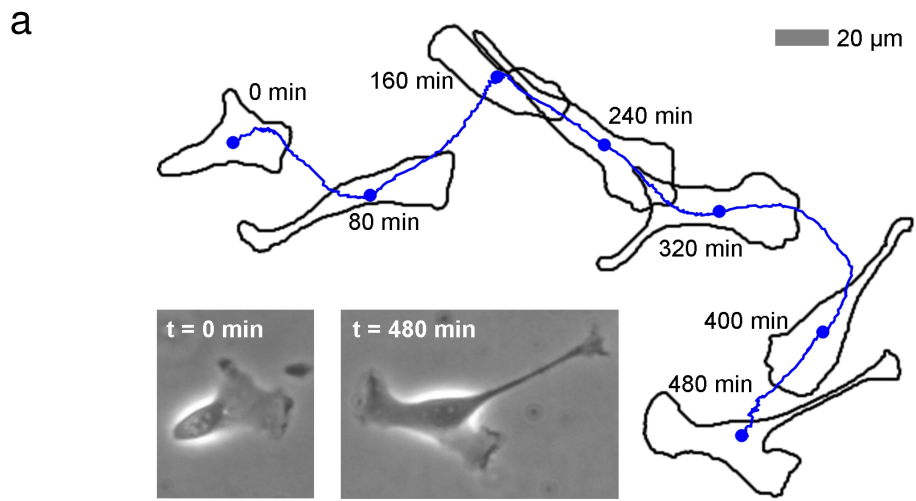
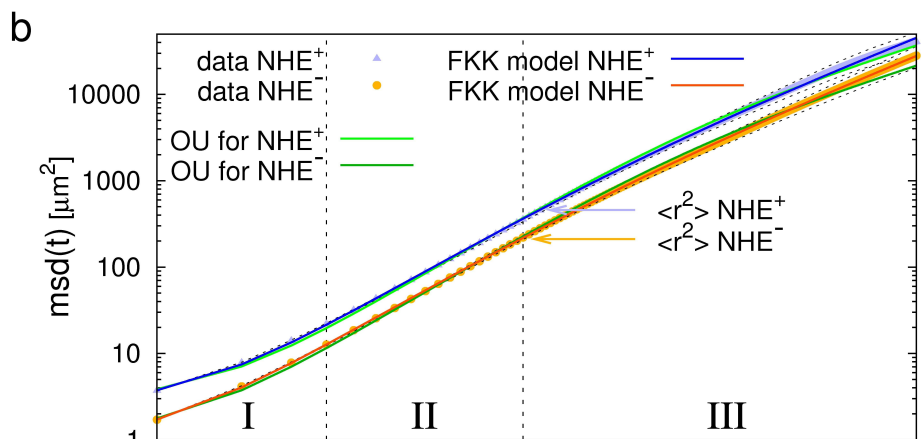
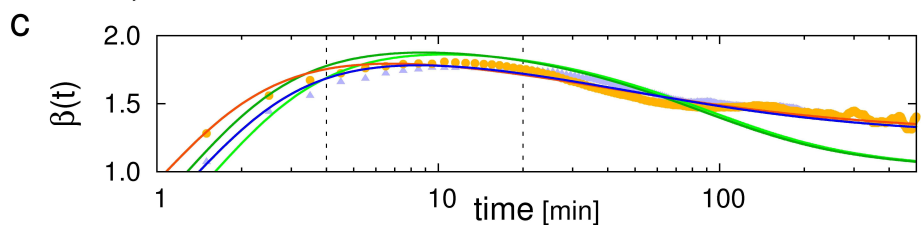

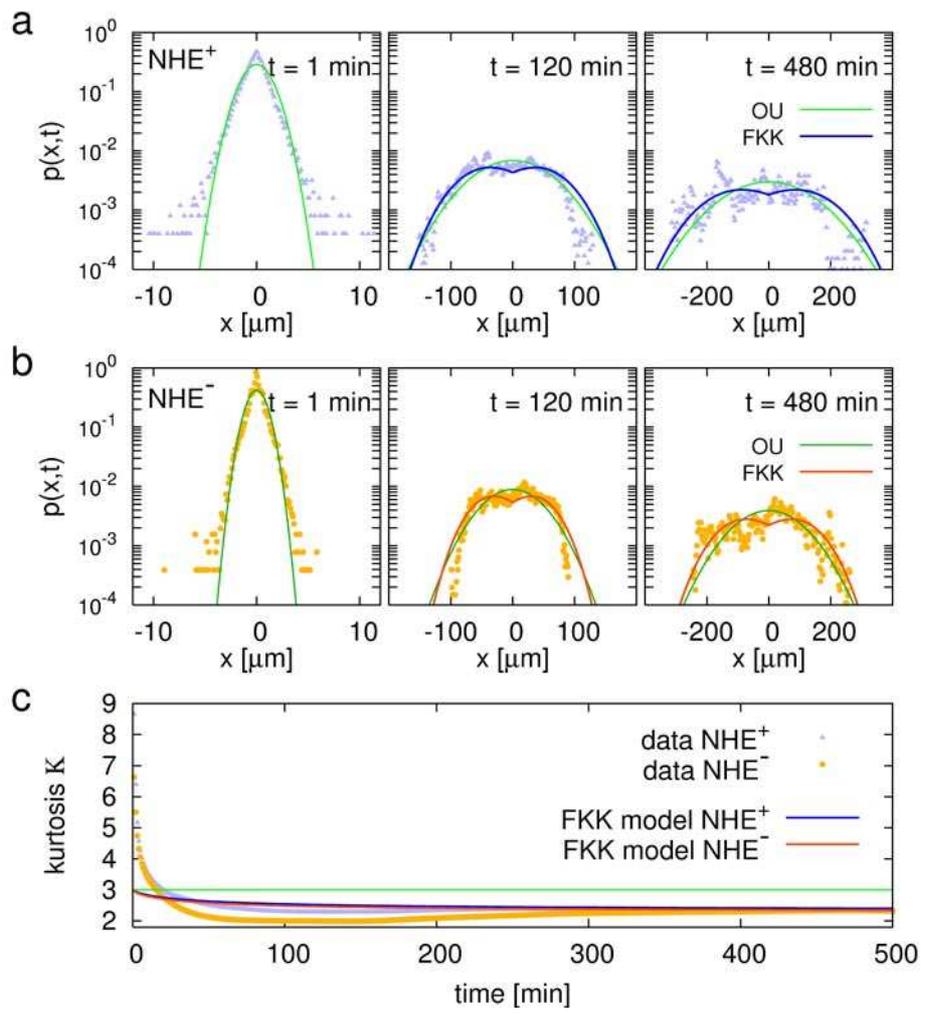

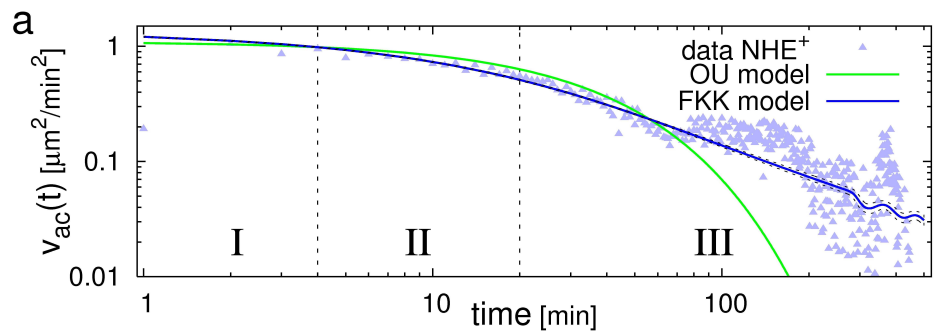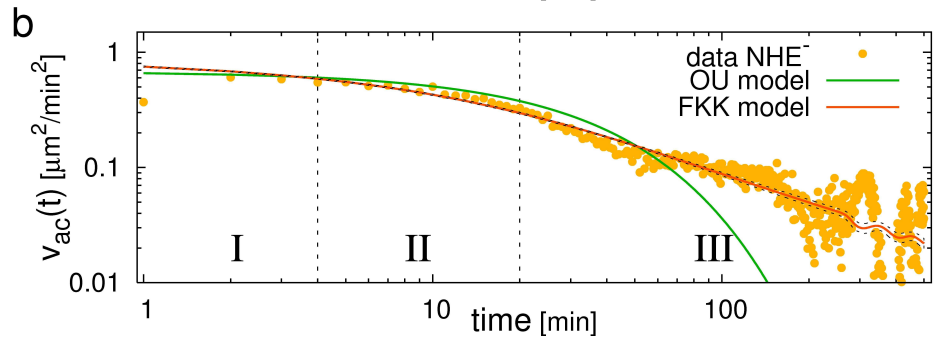